\documentclass{aastex63}

\usepackage[version=4]{mhchem}
\received{}
\revised{}
\accepted{}
\submitjournal{ApJ}

\shorttitle{\ce{H2} ortho-para spin conversion}
\shortauthors{Furuya et al.}

\newcommand{\num}[1]{n({\rm #1})}

\newcommand{\edes}{E_{\rm b}}
\newcommand{\edcop}{E_{\rm b}^{\rm \,op}}
\newcommand{\edcdes}{E_{\rm b}^{\rm \,des}}
\newcommand{\ehop}{E_{\rm hop}}
\newcommand{\phh}{{\rm p\mathchar`- H_{2}}}
\newcommand{\ohh}{{\rm o\mathchar`- H_{2}}}
\newcommand{\pdd}{{\rm p\mathchar`- D_{2}}}
\newcommand{\odd}{{\rm o\mathchar`- D_{2}}}
\newcommand{\op}[1]{{\rm OPR(\ce{#1})}}

\begin{document}

\title{\ce{H2} Ortho-Para Spin Conversion on Inhomogeneous Grain Surfaces}

\correspondingauthor{Kenji Furuya}
\email{furuya@ccs.tsukuba.ac.jp}

\author{Kenji Furuya}
\affiliation{Center for Computer Sciences, University of Tsukuba, 305-8577 Tsukuba, Japan}
\author{Yuri Aikawa}
\affiliation{Department of Astronomy, The University of Tokyo, Bunkyo-ku, Tokyo 113-0033, Japan}
\author{Tetsuya Hama}
\affiliation{Institute of Low Temperature Science, Hokkaido University, Sapporo, Hokkaido 060–0819, Japan}
\author{Naoki Watanabe}
\affiliation{Institute of Low Temperature Science, Hokkaido University, Sapporo, Hokkaido 060–0819, Japan}



\begin{abstract}

We investigate the evolution of the ortho-to-para ratio of overall (gas + ice) \ce{H2} via the nuclear spin conversion 
on grain surfaces coated with water ice
under physical conditions that are relevant to star- and planet-forming regions.
We utilize the rate equation model that considers adsorption of gaseous \ce{H2} on grain surfaces, 
which have a variety of binding sites with a different potential energy depth, thermal hopping, desorption, and the nuclear spin conversion of adsorbed \ce{H2}.
It is found that the spin conversion efficiency depends on the \ce{H2} gas density and the surface temperature. 
As a general trend, enhanced \ce{H2} gas density reduces the efficiency, while the temperature dependence is not monotonic; 
there is a critical surface temperature at which the efficiency is the maximum.
At low temperatures, the exchange of gaseous and icy \ce{H2} is inefficient
(i.e., adsorbed \ce{H2} does not desorb and hinders another gaseous \ce{H2} to be adsorbed),
while at warm temperatures, the residence time of \ce{H2} on surfaces is too short for the spin conversion.
Additionally, the spin conversion becomes more efficient with lowering the activation barriers for thermal hopping.
We discuss whether the spin conversion on surfaces can dominate over that in the gas-phase in star- and planet-forming regions.
Finally, we establish a simple, but accurate way to implement the \ce{H2} spin conversion on grain surfaces in existing gas-ice astrochemical models.

\end{abstract}





\section{Introduction} \label{sec:intro}
Hydrogen is the most abundant element in the universe.
In star- and planet-forming regions, hydrogen is primarily present in \ce{H2}, which has two nuclear spin configurations, ortho and para.
As the internal energy difference between ortho-\ce{H2} and para-\ce{H2} (170.5 K) is much higher than the typical temperature of star-forming regions ($\sim$10 K), 
the ortho-to-para ratio (OPR) of \ce{H2} can significantly affect the molecular evolution, for example, deuterium fractionation \citep[see, e.g.,][]{pagani92,flower06,taquet14,furuya16}. 

\ce{H2} molecules form on grain surfaces with the statistical ortho-to-para ratio of three \citep{watanabe10}.
After the \ce{H2} formation, the ortho-para spin conversion of \ce{H2} proceeds through proton exchange reactions with \ce{H+} and/or with \ce{H3+} in the gas phase \citep{gerlich90,honvault11}. 
Laboratory experiments have found that the \ce{H2} spin conversion can also occur on bare grain \citep[\ce{D2} on graphite surfaces; e.g.,][]{yucel90} and on amorphous water ice surfaces \citep[e.g.,][]{watanabe10} 
in laboratory timescales (around a few hours),
while the mechanism of the spin conversion on the surfaces is not fully understood \citep[see, e.g.,][]{fukutani13,ilisca18}.
Given this very short timescale, it is expected that the spin conversion on surfaces affects the \ce{H2} OPR evolution in star- and planet-forming regions.
However, its efficiency in the astronomical conditions remains unclear for the following two reasons.
First, almost all \ce{H2} is present in the gas phase rather than on grain surfaces.
Then the spin conversion timescale of overall (gas + solid) \ce{H2} via the conversion on surfaces depends on how efficiently gaseous and solid \ce{H2} interact. 
Second, the probability for the nuclear spin state of an adsorbed \ce{H2} molecule to be changed before it is desorbed depends on 
the residence time on surfaces (i.e., thermal desorption timescale) versus the spin conversion timescale.
Interstellar dust grains are coated with ice mantles, the main component of which is water, in the cold ($\lesssim$100 K) gas of star-forming regions \citep[see][for a recent review]{boogert15}.
The surface of the ice mantles would contain various binding sites with a different energy depth.
This is relevant to both points, because in that case, the thermal desorption timescale depends on site. 

\begin{figure}[ht!]
\plotone{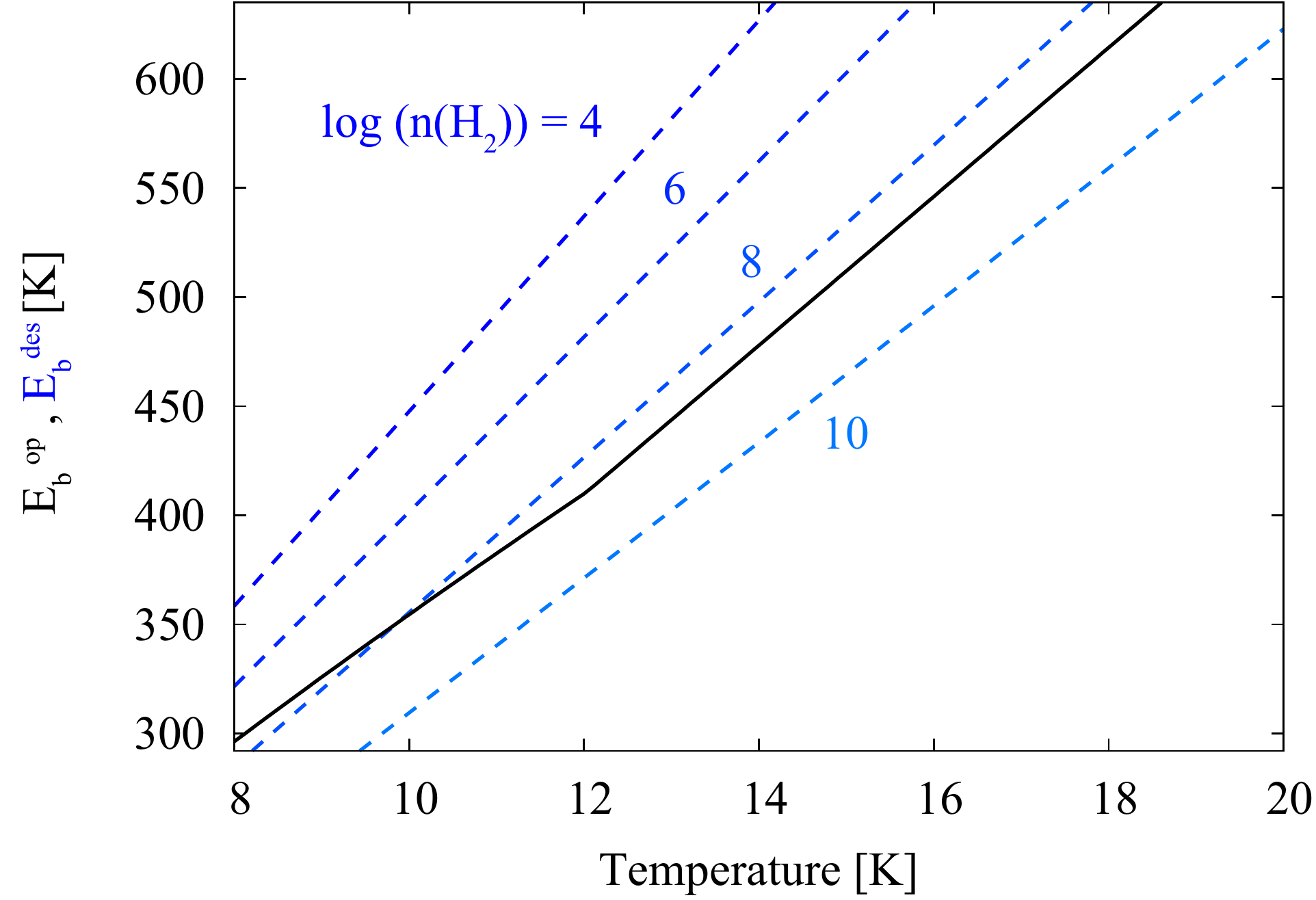}
\caption{$\edcdes$ (dashed blue lines) and $\edcop$ (black solid line) as functions of temperature. 
For $\edcdes$, four different \ce{H2} gas density cases (10$^4$ cm$^{-3}$, 10$^6$ cm$^{-3}$, 10$^8$ cm$^{-3}$, and 10$^{10}$ cm$^{-3}$) are shown. 
$\edcop$ does not depend on the \ce{H2} gas density.}
\label{fig:ecrit}
\end{figure}

In order to see the two points raised above more quantitatively, first, let us consider the balance between the adsorption rate of gaseous \ce{H2} 
on water ice surfaces and the thermal desorption rate of adsorbed \ce{H2};
\begin{align}
\frac{1}{4}(1-\theta(\edes))Sv_{\rm th}n_{\rm site}^{-1}\num{H_2} = \nu \theta(\edes) \exp(-\edes/T), \label{eq:crit_edes}
\end{align}
where $S$ is the sticking probability to the water ice surface, $\num{H_2}$ is the number density of \ce{H2} in the gas phase, $v_{\rm th}$ is the thermal velocity of \ce{H2}, 
$n_{\rm site}$ is the density of binding sites on the surface ($1.5\times10^{15}$ cm$^{-2}$), 
$\nu$ is the vibrational frequency (typically 10$^{12}$ s$^{-1}$), $\edes$ is the binding energy of \ce{H2} on the water ice surface, and $T$ is the temperature of the surface.
$\theta(\edes)$ is the fraction of binding sites occupied by \ce{H2} with the potential energy depth of $\edes$.
We assume that only one \ce{H2} is allowed per binding site, which leads to the factor $1-\theta$ in the left hand side of the equation.  
From this equation, we can define critical binding energy ($\edcdes$) such that all sites with $\edes > \edcdes$ will be occupied by \ce{H2}, 
i.e., $\theta(\edes > \edcdes) = 1$ \citep{dissly94}.
At $T = 10 $ K and $\num{H_2} = 10^4$ cm$^{-3}$, $\edcdes$ is 440 K.
Let us define another critical binding energy ($\edcop$) such that the thermal desorption timescale of \ce{H2} 
in binding sites with $\edes > \edcop$ is long enough for the conversion of ortho-\ce{H2} to para-\ce{H2}.
By considering the balance between the spin conversion rate of ortho-\ce{H2} to para-\ce{H2} ($k_{\rm op}^{\rm surf}$) and the thermal desorption rate ($\nu \exp(-\edes/T)$), 
we obtain $\edcop$ of $\sim$360 K at the surface temperature of 10 K and for $k_{\rm op}^{\rm surf}$ of $3\times10^{-4}$ s$^{-1}$ \citep{ueta16}.
Based on theses arguments, one may think that binding sites which satisfies $\edcop \lesssim \edes \lesssim \edcdes$ contribute to the evolution of the \ce{H2} OPR most efficiently;
for $\edes < \edcop$, the residence time is too short for the spin conversion, 
while for $\edcdes < \edes$, adsorbed \ce{H2} does not desorb efficiently and hinders another gaseous \ce{H2} to be adsorbed.
Then binding energy distribution does matter, and the question is what fraction of sites have binding energy in the range of  $\edcop \lesssim \edes \lesssim \edcdes$.
Note that $\edcop$, $\edcdes$, and their inequality relation depend on physical conditions as shown in Figure \ref{fig:ecrit}.
The above discussion, however, neglects thermal hopping of adsorbed \ce{H2}.
As we will see later, thermal hopping changes the situation significantly, because it allows adsorbed \ce{H2} molecules to visit various sites with a different potential energy depth.
In summary, to understand the spin conversion efficiency of \ce{H2} in the astronomical conditions, 
one has to consider \ce{H2} adsorption on the surface, which contains a variety of sites, thermal desorption and hopping, and the nuclear spin conversion in a self-consistent way.
In this work, we construct such a model for the first time.

\epsscale{0.7}
\epsscale{1.0}

The effect of the spin conversion on the surface on the \ce{H2} OPR evolution was theoretically studied by \citet{bron16} in the context of photodissociation regions (PDRs) 
and by \citet{bovino17} in the context of dense molecular clouds.
Both models did not consider the binding energy distribution of \ce{H2}, but used a single ``representative'' binding energy as commonly assumed in astrochemical models for simplicity. 
\citet{bron16} found that the fluctuation of dust temperature due to stochastic heating by UV photons is important for determining the spin conversion efficiency in PDRs.
\citet{bovino17} discussed $\edcop$ considering uncertainties of  relevant parameters, but their discussion lacks another necessary condition, $\edcdes$.

This paper is organized as follows: 
our numerical model is described in Sect. \ref{sec:model} 
and the results are discussed in in Sect. \ref{sec:result}.
In Sect. \ref{sec:discuss}, we propose a simple model that reproduces our numerical results, 
and discuss whether the spin conversion on surfaces can dominate over that in the gas phase in star- and planet-forming regions.
Our findings are summarized in Sect. \ref{sec:conclusion}.

\section{Methods}
\label{sec:model}
\subsection{Basic equations}
We adopt a rate equation approach to investigate the efficiency of the ortho-para conversion on grain surfaces in star and planet forming regions.
We consider a typical interstellar grain with radius of 0.1 $\mu$m with the dust-to-gas mass ratio of $10^{-2}$.
The grain is assumed to be covered by water ice mantles and the number of binding sites on the water ice surface per area ($n_{\rm site}$) is set to be $1.5\times10^{15}$ cm$^{-2}$.
The total number of binding sites per grain is thus $N_{\rm site} \approx 2 \times 10^6$.
We consider the experimental fact that the water ice surface contains various sites with a different potential energy depth \citep[e.g.,][]{amiaud06}.
For simplicity, we assume the following throughout this work: 
(1) only one molecule is allowed to be adsorbed per binding site, 
(2) ortho-\ce{H2} ($\ohh$) and para-\ce{H2} ($\phh$) share common binding sites, following the same binding energy distribution (see Section \ref{sec:binding}), and
(3) chemical properties of $\ohh$ and $\phh$ are the same except that they convert to each other on the surface with different rates.
We denote the fraction of biding sites, which are occupied by $\ohh$ ($\phh$) as $\theta_o$ ($\theta_p$). 
The following condition should be satisfied:
\begin{equation}
\theta(\edes, t) = \theta_o(\edes, t) + \theta_p(\edes, t).
\end{equation}
We denote the binding energy distribution of \ce{H2} on the surface as $g$, which satisfies
\begin{equation}
\int g(\edes)d\edes = 1.
\end{equation}
The surface coverage of \ce{H2} at a given time $t$, $\Theta(t)$, is defined as
\begin{equation}
\Theta(t) = \int \theta(\edes, t)g(\edes)d\edes. \label{eq:Thetat}
\end{equation}
Similarly $\Theta_{\alpha}(t)$, where $\alpha$ is $o$ or $p$, is defined as
\begin{equation}
\Theta_{\alpha}(t) = \int \theta_{\alpha}(\edes, t)g(\edes)d\edes, 
\end{equation}
and thus $\Theta(t) = \Theta_{o}(t) + \Theta_{p}(t)$.

We numerically solve the following rate equations, which describe adsorption of \ce{H2}, thermal desorption, thermal hopping, 
and spin conversion of adsorbed \ce{H2}, considering various binding sites with a different potential energy depth \citep[cf.][]{li10}:
\begin{align}
\frac{d\num{\alpha \mathchar`- \ce{H2}}}{dt} &= - (1-\Theta(t))SR_{\rm col}(\alpha \mathchar`- \ce{H2}) + R_{\rm thdes}(\alpha \mathchar`- \ce{H2}), \label{eq:gas_h2} \\
\frac{d\theta_{\alpha}(\edes, t)}{dt} &= \frac{1}{4}[1-\theta(\edes, t)]Sv_{\rm th} n_{\rm site}^{-1}\num{\alpha \mathchar`- \ce{H2}} - k_{\rm thdes}(\edes)\theta_{\alpha}(\edes, t)  \label{eq:cov_h2} \\ 
&- \int k_{\rm hop}(\edes \rightarrow \edes') \theta_{\alpha}(\edes, t) [1-\theta(\edes', t)]g(\edes')d\edes' \nonumber \\
&+  [1-\theta(\edes, t)]\int k_{\rm hop}(\edes' \rightarrow \edes) \theta_{\alpha}(\edes', t)g(\edes')d\edes' \nonumber \\
&+ k^{surf}_{\beta\alpha}\theta_{\beta}(\edes, t) - k^{surf}_{\alpha\beta}\theta_{\alpha}(\edes, t), \nonumber 
\end{align}
where $\alpha$ and $\beta$ indicate either ortho ($o$) or para ($p$).
The collision rates to dust grains and desorption rates from the whole surface of dust grains of  $\ohh$ and $\phh$ are given by
\begin{align}
R_{\rm col}(\alpha \mathchar`- \ce{H2}) &= v_{\rm th} \sigma \num{\alpha \mathchar`- \ce{H2}}n_{\rm gr}, \\
R_{\rm thdes}(\alpha \mathchar`- \ce{H2}) &= n_{\rm gr}N_{\rm site}\int k_{\rm thdes}(\edes')\theta_{\alpha}(\edes', t)g(\edes')d\edes',
\end{align}
where $v_{\rm th}$ is the thermal velocity, $\sigma$ is the cross section of a dust grain, $n_{\rm gr}$ is the number density of dust grains per unit gas volume,
and $k_{\rm thdes}$ is the thermal desorption rate (s$^{-1}$).
We assume gas and surface temperatures are the same and do not distinguish them throughout this paper.

The first terms in Eqs. \ref{eq:gas_h2} and \ref{eq:cov_h2} represent adsorption with the sticking probability $S$ of \ce{H2} to the water ice surface.
We consider the factor $1-\Theta$ or the factor $1-\theta$, because only one molecule is allowed to be adsorbed per binding site in our models.
Then the maximum value of $\Theta$ is unity and the formation of \ce{H2} multilayers does not occur in our models.
Indeed, laboratory experiments have found that no matter how large \ce{H2} fluence deposited on a water ice substrate is, 
the \ce{H2} coverage is in the submonolayer regime even at 10 K \citep[e.g.,][]{gavilan12,kuwahata15}.
The second terms in Eqs. \ref{eq:gas_h2} and \ref{eq:cov_h2} represent thermal desorption.
The third term in Eq. \ref{eq:cov_h2} represents thermal hopping from a binding site with $\edes$ to a site with $\edes'$, while the forth term represents the reverse process.
The hopping activation energy in our models is discussed later.
The fifth and sixth terms are for ortho-para conversion on surfaces, the rates of which are discussed in Section \ref{sec:tau_conv}.
Initially, all \ce{H2} are assumed to be present in the gas phase with the $\op{H2}$ of three (i.e., the statistical value).

\subsection{Binding energy distribution, hopping activation energy, and sticking probability}
\label{sec:binding}
We use binding energy distribution and sticking probability of \ce{H2} that are appropriate for nonporous amorphous solid water (ASW) in this work.
The degree of porosity of interstellar ice, which is mainly composed of water, remains unclear.
There is no clear observational evidence that interstellar water ice has a porous structure;
the OH dangling bonds of water ice have not been detected in the midinfrared spectrum in the interstellar matter \citep[ISM;][]{keane01},
although the non-detection might be due to the sensitivity limitations of the Infrared Space Observatory (ISO).
\citet{oba09} found that in their experiments, water ices formed from atomic hydrogen and molecular oxygen at low temperatures (10 K-40 K) present a nonporous structure 
compared to vapor deposited water ices at the low temperatures.   
\citet{garrod13} found that in their off-lattice Monte-Carlo simulations, ices formed by surface chemistry under dark cloud conditions present a nonporous structure, 
being consistent with the experiments.
In addition, laboratory experiments have found that the porosity of amorphous water ice decreases after UV photon irradiation and/or 
cosmic-ray impacts \citep[e.g.,][]{raut08,palumbo10}. 
Taken together, nonporous ASW could be more representative for interstellar water ice rather than porous ASW.

The thermal desorption rate depends on the binding energy of the species to the surface,
\begin{equation}
k_{\rm thdes} = \nu \exp(-\edes/T).
\end{equation}
Our binding energy distribution of \ce{H2}, which ranges from 290 K to 635 K, is divided into 100 equal intervals in our simulations (see black line in the top panel of Figure \ref{fig:h2op_cov}).
The binding energy distribution of \ce{D2} on nonporous ASW is available in the literature, 
which was obtained from direct inversion of temperature programmed desorption spectra \citep{amiaud07,he14}.
We obtained the binding energy distribution of \ce{H2} by considering zero-point energy difference between \ce{D2} and \ce{H2}, 3.15 meV \citep{amiaud15}.
Note that laboratory experiments have found that $\odd$ is bound to surfaces slightly more strongly than $\pdd$ ($\sim$1 meV)  \citep[][]{amiaud08,tsuge19}, but
we neglect the difference in this work for simplicity.

The hopping activation energy from a site with the binding energy of $\edes$ to another site with the binding energy of $\edes'$ ($\ehop$) is given as follows \citep[][see their Fig. 11]{cazaux17}:
\begin{equation}
\ehop(\edes \rightarrow \edes') = f \times {\rm min}(\edes,\,\,\edes') + {\rm max}(0,\,\,\edes - \edes'), \label{eq:ehop}
\end{equation}
where $f$ is a free parameter.
The parameter $f$, which is the hopping-to-binding energy ratio, is poorly constrained and values between 0.3 and 0.8 are normally assumed in the astrochemical community.
We choose $f = 0.5$ in our fiducial model. 
Given the expression of $\ehop$, the thermal hopping rate, $k_{\rm hop} = \nu \exp(-\ehop/T)$, obeys the microscopic reversibility, 
i.e., $k_{\rm hop}(\edes \rightarrow \edes')/k_{\rm hop}(\edes' \rightarrow \edes) = \exp[-(\edes-\edes')/T]$  \citep{cuppen13}.

\citet{he16} experimentally investigated the sticking probability for stable molecules on nonporous ASW in low surface coverage regime (below 10 \%).
For sticking probability of \ce{H2} onto water ice surfaces ($S$), we use the formula recommended by \citet{he16} (see their Eq. 1).
For example, $S$ is $\sim$0.7 at 10 K and $\sim$0.5 at 16 K. 
The experimental values may be considered as the surface averaged value, while the sticking probability for each site may depend on the energy depth of each site.
Such (possible) complexity is not considered in our models, i.e, $S$ is set to be the same for all binding sites. 

\subsection{Ortho-para conversion rates on surfaces}
\label{sec:tau_conv}
The ortho-para conversion timescale of \ce{H2} on amorphous water ice ($\tau^{\rm surf}_{\rm conv}$) in the temperature range between 9 K and 16 K was measured in laboratory by \citet{ueta16}.
From $\tau^{\rm surf}_{\rm conv}$, the rate of the conversion 
from $\ohh$ to $\phh$ ($k_{\rm op}^{\rm surf}$) and that of the reverse process ($k_{\rm po}^{\rm surf}$) can be deduced to be 
\begin{align}
k_{\rm op}^{\rm surf} &= (\tau^{\rm surf}_{\rm conv}(1 + \gamma))^{-1}, \\ 
k_{\rm po}^{\rm surf} &= k_{\rm op}^{\rm surf}\gamma,
\end{align}
where $\gamma$ is the thermalized value of $\op{H2}$, $9\exp(-170.5/T)$ \citep{bron16}, 
assuming the energy difference between $\ohh$ and $\phh$ on water ice surfaces is the same as that in the gas phase.
On water ice surfaces, \ce{H2} molecules would not rotate freely and 
thus the energy difference between $\ohh$ and $\phh$ would become smaller than that in the gas phase, but the exact value remains unclear
\citep[cf. see][for the discussion on the energy difference between ortho-\ce{H2O} and para-\ce{H2O} on surfaces]{hama16}.
\citet{ueta16} found that at the temperature lower than $\sim$12 K, $\tau_{\rm conv}$ is fitted by a power low $1/(AT^n)$, where $A$ is $3.2\times10^{-11}$ s$^{-1}$ and $n$ is 7.1.
At the higher temperature, $\tau^{\rm surf}_{\rm conv}$ is almost constant with the value of around $1/(1.5\times10^{-3}) \approx 670$ s.
We take  $\tau^{\rm surf}_{\rm conv}$ from \citet{ueta16} with the lower limit of  670 s.
At 10 K, for example, $k_{\rm op}^{\rm surf}$ and $k_{\rm po}^{\rm surf}$ are $3.1\times10^{-4}$ s$^{-1}$ and $10^{-12}$ s$^{-1}$, respectively.

\section{Results}
\label{sec:result}
\subsection{\ce{H2} coverage}
\begin{figure}[ht!]
\plotone{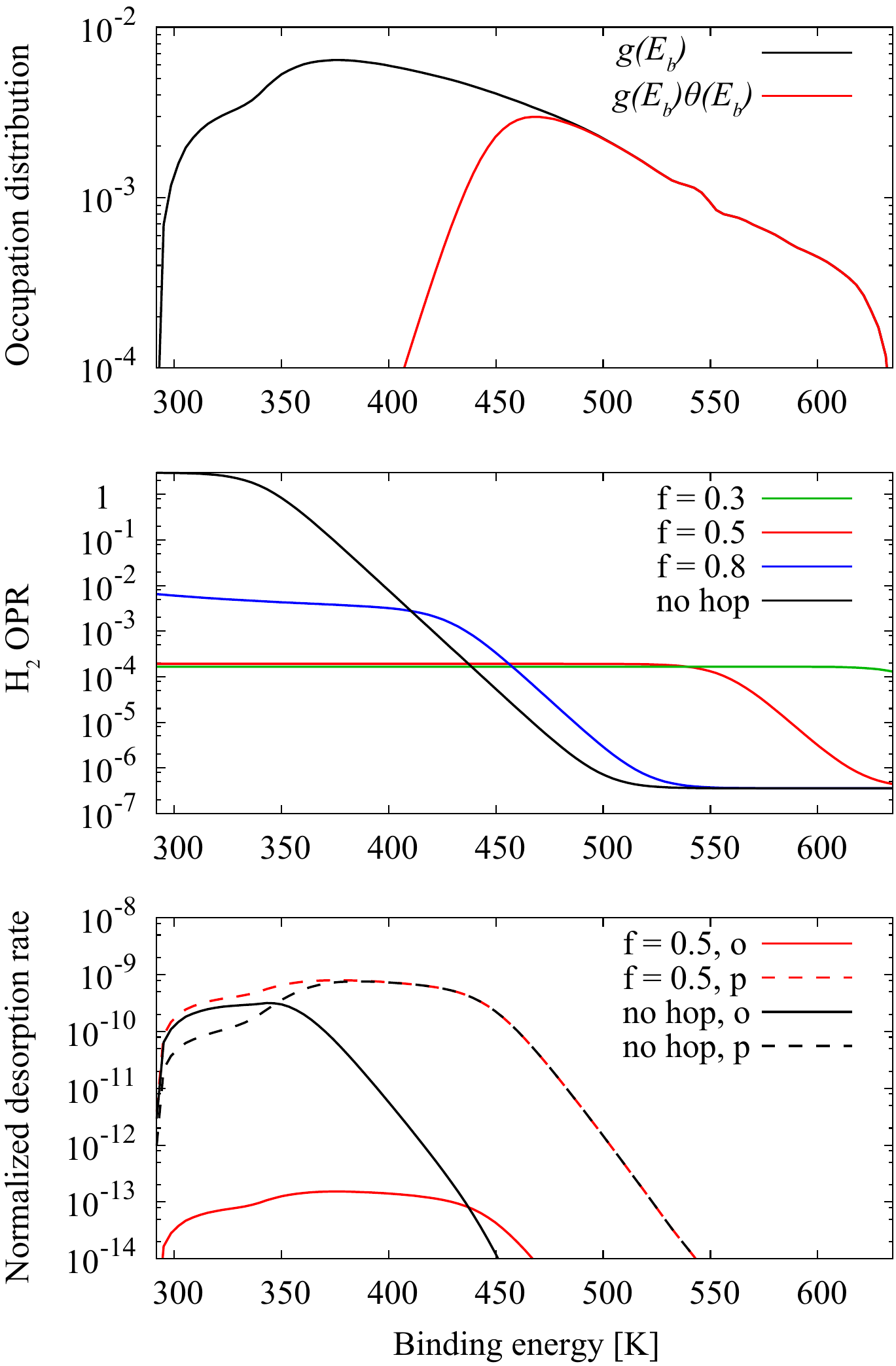}
\caption{Top):Occupation distribution of the binding sites at the adsorption-desorption equilibrium  at $\num{H_2} = 10^4$ cm$^{-3}$ and $T = 10$ K (red line). 
Black solid line shows the binding energy distribution on the whole surface taken from \citet{he14}, but shifts 3.15 meV to 
the lower energy side \citep{amiaud15}.
Middle): $\op{H2}$ ratio on the surface as functions of $\edes$  at $\num{H_2} = 10^4$ cm$^{-3}$ and $T = 10$ K.
Color lines show the model with thermal hopping, varying the parameter $f$, 0.3 (green), 0.5 (red), and 0.8 (blue). 
Black line shows the model without thermal hopping.
Bottom): Normalized desorption rates of $\ohh$ (solid lines) and $\phh$ (dashed lines) from sites with the potential energy depth of $\edes$.
Red lines show the model with $f = 0.5$, while black lines show the model without thermal hopping.}
\label{fig:h2op_cov}
\end{figure}

\begin{figure}[ht!]
\plotone{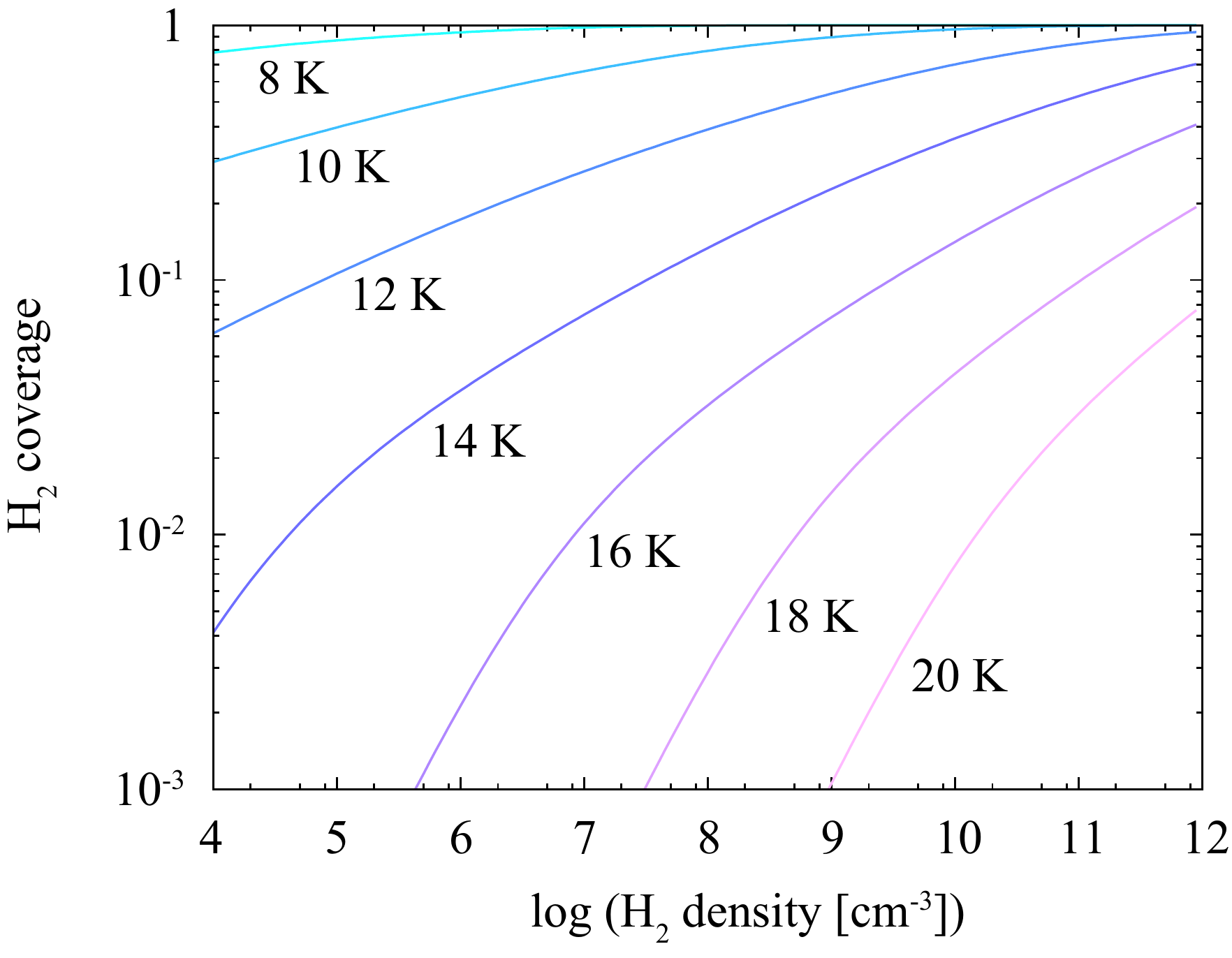}
\caption{Equilibrium \ce{H2} coverage on the water ice surface as functions of \ce{H2} density in the gas phase, varying temperature from 8 K to 20 K.
}
\label{fig:cov_eq}
\end{figure}

The \ce{H2} coverage on the water ice surface is discussed in detail in our separate work \citep{furuya19}, 
where the similar rate equations to Eqs. \ref{eq:gas_h2} and \ref{eq:cov_h2} are used, but without distinction of the \ce{H2} nuclear spin states.
We briefly summarize this here. 
The adsorption and desorption of \ce{H2} reach the equilibrium in a very short timescale ($\lesssim$1 yr ($10^4$ cm$^{-3}$/$\num{H_2}$)). 
Then only the equilibrium condition is relevant in the dense ISM. 
The occupation of sites with the potential energy depth of $\edes$ is then determined by the balance 
between the adsorption rate of gaseous \ce{H2} on each site and the thermal desorption rate of adsorbed \ce{H2}, 
and one can obtain $\theta(\edes)$ by solving Eq. \ref{eq:crit_edes} \citep[see also][]{amiaud06};
\begin{align}
\theta(\edes) &= \left(1 + \exp\left(-\frac{\edes-\edcdes}{T}\right)\right)^{-1}, \label{eq:fd_dist} \\
\edcdes &= T \ln \left(\frac{4\nu n_{\rm site}}{Sn(\ce{H2})v_{\rm th}}\right). \label{eq:edcdes}
\end{align}
$\edcdes$ is the critical binding energy such that a half of sites with $\edes$ will be occupied by \ce{H2}, i.e., $\theta(\edcdes) = 0.5$, under the adsorption-desorption equilibrium.
Thus all sites with $\edes \gg \edcdes$ will be occupied by \ce{H2}, $\theta(\edes \gg \edcdes) = 1$.
As an example, the top panel of Figure \ref{fig:h2op_cov} shows the equilibrium occupation distribution ($\theta \times g$) of \ce{H2} at $\num{H_2} = 10^4$ cm$^{-3}$ and $T = 10$ K, 
where the \ce{H2} surface coverage ($\Theta$) is $\sim$30 \%.
It shows that deeper sites are preferentially occupied by \ce{H2}.
The equilibrium \ce{H2} coverage as functions of the \ce{H2} gas density and temperature are shown in Figure \ref{fig:cov_eq}.
The \ce{H2} coverage increases with increasing the gas density and with decreasing the temperature.
The occupation distribution and the \ce{H2} surface coverage at the equilibrium do not depend on the hopping parameter $f$.

\epsscale{0.7}
\epsscale{1.0}

\epsscale{0.7}
\epsscale{1.0}

\subsection{\ce{H2} ortho-para spin conversion in the fiducial physical conditions}
We first show model results at $\num{H_2} = 10^4$ cm$^{-3}$ and $T = 10$ K (our fiducial physical conditions) 
and discuss the dependence on the physical conditions later.
The middle panel of Figure \ref{fig:h2op_cov} shows the $\op{H2}$ on the surface as functions of $\edes$ (i.e., $\theta_o/\theta_p$ for each $\edes$), 
varying  the parameter $f$.
We chose the time when the \ce{H2} fluence (the time integral of the \ce{H2} flux) reaches $5\times10^{16}$ cm$^{-2}$, corresponding to the duration time of $\sim10 (10^4$ cm$^{-3}$/$\num{H_2}$) yr.
By that time, the \ce{H2} coverage on the surface reaches the adsorption-desorption equilibrium at all physical conditions explored in this work, 
while the duration time is too short for the spin conversion of the overall (gas + solid) $\op{H2}$.
Then the $\op{H2}$ in the gas phase remains unchanged from the initial value of three.
The chosen duration time is shorter than $\tau^{\rm surf}_{\rm conv}$ for $\num{H_2} \geq 10^9$ cm$^{-3}$; then we choose $t\sim$10$^{-3}$ yr ($\gg \tau^{\rm surf}_{\rm conv}$) at the higher densities, 
corresponding to the \ce{H2} fluence of $5\times10^{16}$ ($\num{H_2}/10^8$ cm$^{-3}$) cm$^{-2}$.

When the thermal hopping is turned off, $\theta_o/\theta_p$ is determined 
by the timescale of thermal desorption from the site ($k_{\rm thdes}^{-1}$) versus the spin conversion timescale; 
sites with higher $\edes$ have lower $\theta_o/\theta_p$ due to the longer thermal desorption timescale (i.e., the longer resident timescale).
When the thermal hopping is turned on, the situation changes; adsorbed \ce{H2} can visit multiple sites via thermal hopping.
In the fast hopping cases ($f \leq 0.5$), $\theta_o/\theta_p$ is almost constant across the surface.
This indicates that the resident time of \ce{H2} on the surface is independent of the energy depth of  a site 
in which a \ce{H2} molecule was initially adsorbed, 
due to the efficient thermal hopping after adsorption on the surface.
Then $k_{\rm thdes}^{-1}$ is not a good measure of the resident time of adsorbed \ce{H2}, when thermal hopping is considered.
While $\theta_o/\theta_p$ for given $\edes$ is very different depending on the hopping rate,
the OPR averaged on the whole surface ($\Theta_o/\Theta_p$) is similar regardless of the hopping rate; 
e.g., $\Theta_o/\Theta_p = 1.7\times10^{-4}$ in the model with $f = 0.5$, while it is $1.2\times10^{-4}$ in the model without thermal hopping.

In terms of the $\op{H2}$ evolution in the ISM, the ortho-para ratio of desorbing gas (i.e., $R_{\rm thdes}(\ohh)/R_{\rm thdes}(\phh)$) 
is more relevant than that of the surface ($\Theta_o/\Theta_p$), 
as almost all \ce{H2} is present in the gas phase rather than on the surface.
We find that relation between $R_{\rm thdes}(\ohh)/R_{\rm thdes}(\phh)$ and $\Theta_o/\Theta_p$ depends on the efficiency of thermal hopping;
they are similar in the models with fast hopping ($1.9\times10^{-4}$ versus $1.7\times10^{-4}$ for $f=0.5$), 
while they are very different in the model without thermal hopping ($2.7\times10^{-1}$ versus $1.2\times10^{-4}$).
Thermal desorption rates of $\ohh$ and $\phh$ as functions of $\edes$ are shown in the bottom panel of Figure \ref{fig:h2op_cov}.
\ce{H2} desorption predominantly occurs from sites with $\edes \lesssim \edcdes$ (by definition).
In the model without thermal hopping, only sites with $\edcop \lesssim \edes \lesssim \edcdes$ contribute to the decrease of $R_{\rm thdes}(\ohh)/R_{\rm thdes}(\phh)$;
for $\edes < \edcop$, the residence time is too short for the spin conversion, 
while for $\edcdes < \edes$, adsorbed \ce{H2} does not desorb efficiently.
On the other hand, in the model with thermal hopping, sites with $\edes \gtrsim \edcdes$ also contribute to the decrease of $R_{\rm thdes}(\ohh)/R_{\rm thdes}(\phh)$; 
they trap \ce{H2} molecules, the spin states of the \ce{H2} molecules are converted, and after some time, the \ce{H2} molecules hop to shallower sites and desorb to the gas phase.
These results demonstrate that the binding energy distribution and the thermal hopping among sites are essentially important 
for the spin conversion on grain surfaces in the ISM.

\begin{figure}[ht!]
\plotone{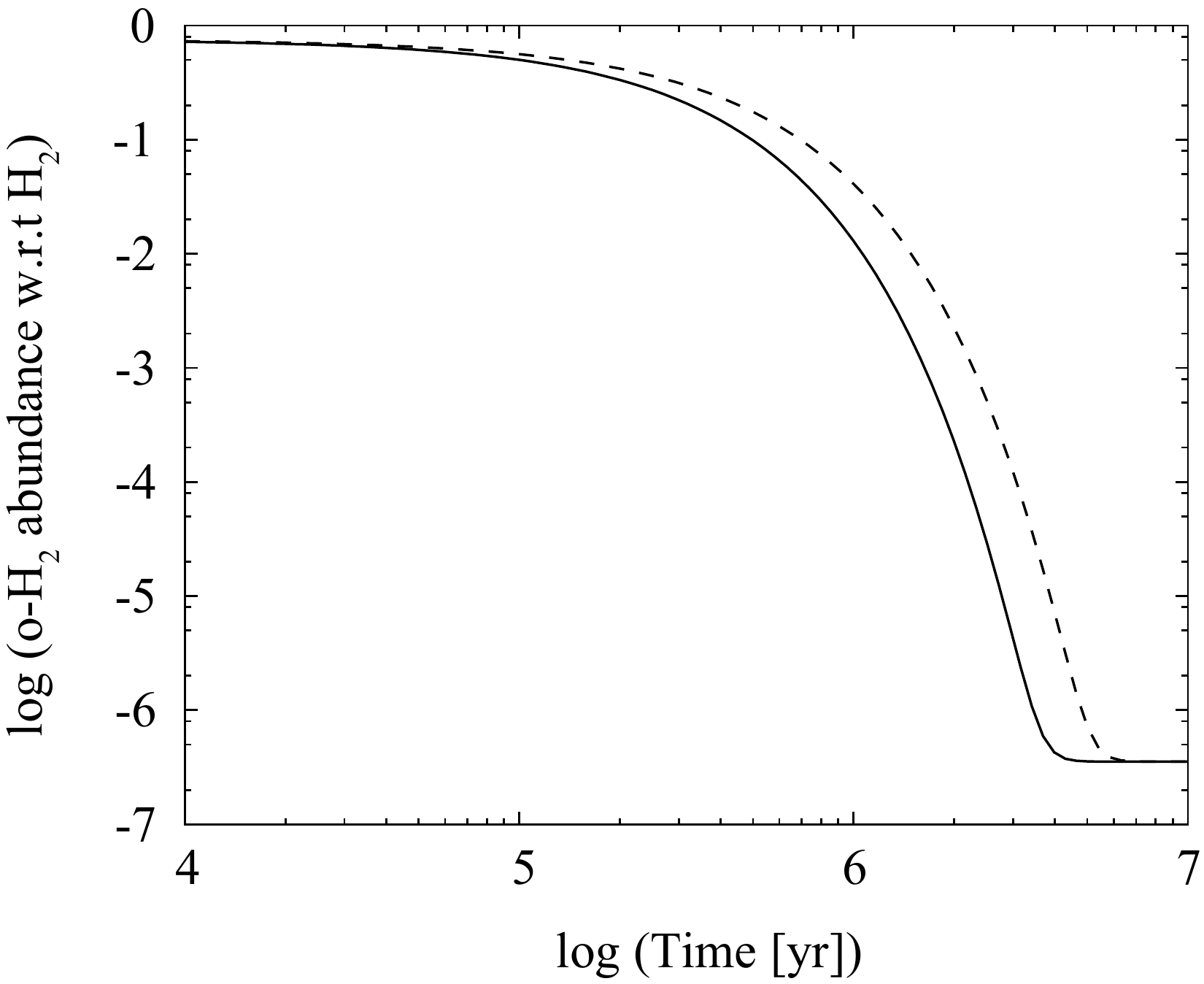}
\caption{Temporal evolution of the abundance of $\ohh$ in the gas phase with respect to \ce{H2} in the model with $f = 0.5$ (solid line) and without thermal hopping (dashed line).
The physical conditions are fixed to $\num{H_2} = 10^4$ cm$^{-3}$ and $T = 10$ K.} 
\label{fig:op_1e4}
\end{figure}

Figure \ref{fig:op_1e4} shows the long term evolution of the $\ohh$ abundance in the gas phase with respect to \ce{H2} at $\num{H_2} = 10^4$ cm$^{-3}$ and $T = 10$ K.
The gaseous $\ohh$ abundance decreases with time due to the spin conversion on the surface.
The spin conversion timescale of $\ohh$ to $\phh$ is in the order of 10$^5$ yr, and the timescale is shorter in the model with $f=0.5$ than that in the model without hopping by a factor of $\lesssim$2.
The spin conversion timescale of gaseous $\ohh$ is given by 
\begin{equation}
\tau_{op} = \num{\ohh}/[R_{\rm thdes}(\phh)-R_{\rm ads}(\phh)],
\end{equation}
where $R_{\rm ads}$ is the adsorption rate of \ce{H2} on dust grain surfaces ($= (1-\Theta)SR_{\rm col}$).
We confirmed that $x_0(\ohh)\exp(-t/\tau_{op})$, where $x_0(\ohh)$ is the initial abundance of gaseous $\ohh$ with respect to \ce{H2}, 
reproduces the numerical results shown in Fig. \ref{fig:op_1e4}.
Note that if the spin conversion on the surface does not occur, $R_{\rm thdes}(\phh) = R_{\rm ads}(\phh)$ under the adsorption-desorption equilibrium.
The steady-state abundance is given by $9\exp(-170.5/10) \approx 3\times10^{-7}$.

\epsscale{0.7}
\epsscale{1.0}

\subsection{Density dependence}
\begin{figure}[ht!]
\plotone{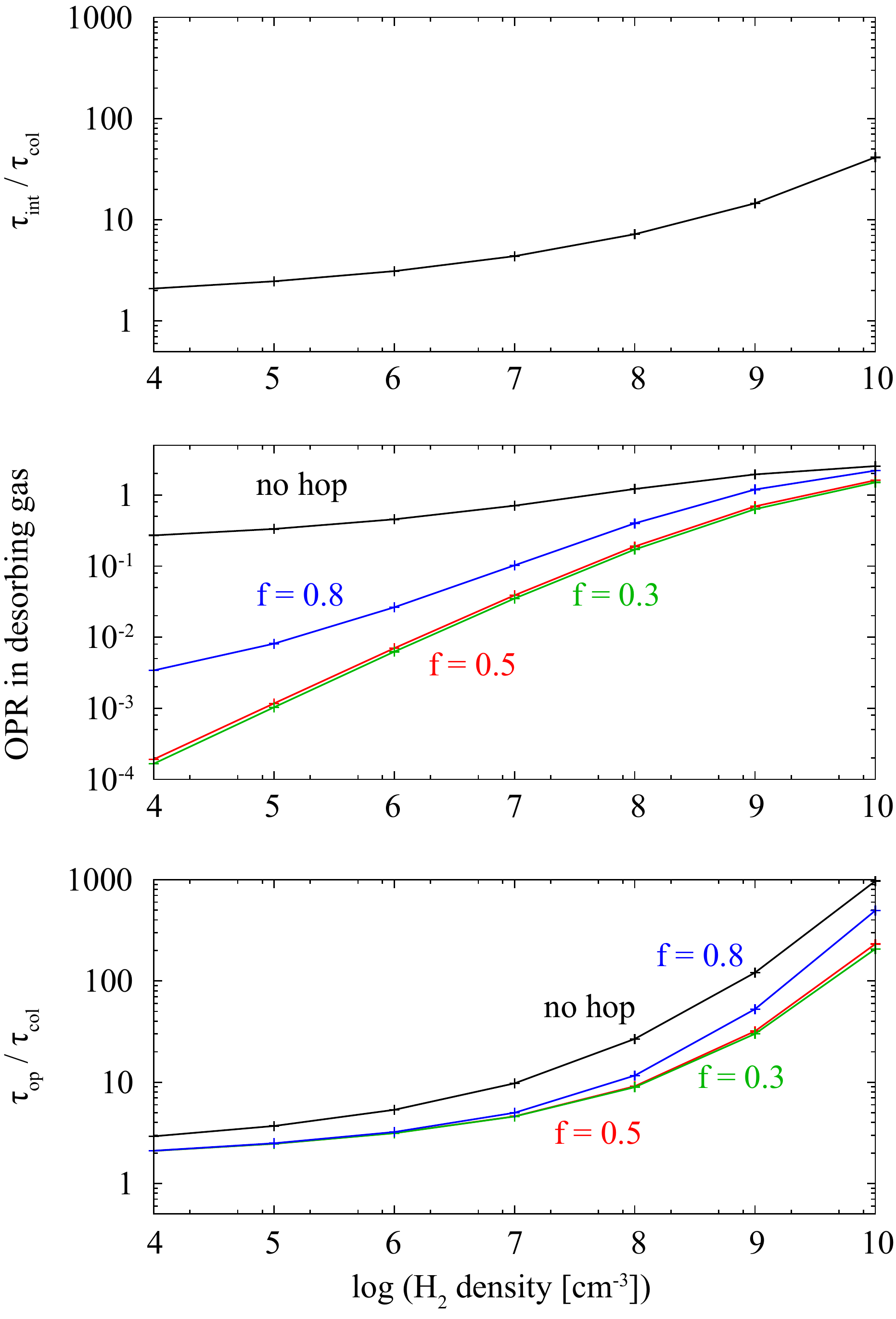}
\caption{Timescale of gaseous and solid \ce{H2} interaction normalized by the collision timescale ($\tau_{int}/\tau_{\rm col}$, top panel), 
\op{H2} in the desorbing gas (middle panel), and 
the spin conversion timescale of $\ohh$ in the gas phase normalized by the collision timescale ($\tau_{op}/\tau_{\rm col}$, bottom panel) as function of \ce{H2} gas density. 
Note that $\tau_{\rm col}$ is inversely proportional to $\num{H_2}$ and thus $\tau_{\rm op}$ in fact drops with increasing $\num{H_2}$.
Temperature is fixed to be 10 K.
The values at $t \sim10 (10^4$ cm$^{-3}$/$\num{H_2}$) yr or $t \sim 10^{-3}$ yr, whichever is longer, are shown.
}
\label{fig:op_dens}
\end{figure}

Here we discuss the density dependence of the spin conversion timescale of gaseous $\ohh$ via the spin conversion on the surface ($\tau_{op}$).
Again, we focus on the results at $t \sim10 (10^4$ cm$^{-3}$/$\num{H_2}$) yr for $\num{H_2} \leq 10^8$ cm$^{-3}$, 
while at higher densities, we focus on the results at $t \sim10^{-3}$ yr.
As discussed in the Introduction, there are two main factors that control $\tau_{op}$:
(i) the efficiency of the interaction between gaseous and solid \ce{H2} and 
(ii) the probability of the spin-state conversion of an adsorbed \ce{H2} molecule before it is desorbed.
The top panel of Figure \ref{fig:op_dens} shows the timescale for gaseous and solid \ce{H2} interaction defined 
by $\tau_{\rm int} = \num{H_2}/R_{\rm thdes}(\ce{H2})$ (or equivalently $\num{H_2}/R_{\rm ads}(\ce{H2})$ at the adsorption-desorption equilibrium) as function of the \ce{H2} gas density.
For convenience, we normalize $\tau_{\rm int}$ by the collisional timescale, $\tau_{\rm col} = \num{H_2}/R_{\rm col}(\ce{H2}) = (v_{\rm th} \sigma n_{\rm gr})^{-1} \approx 3\times10^9/\num{\ce{H2}}$ yr.
The normalized interaction timescale ($\tau_{\rm int}/\tau_{\rm col} = 1/[S(1-\Theta)]$), which means the average number of collisions for an \ce{H2} molecule required to be adsorbed on the water ice surface, 
becomes larger with increasing  $\num{H_2}$, 
because  $\edcdes$ becomes smaller (see Fig. \ref{fig:ecrit}) and thus $\Theta$ increases.

The middle panel of Figure \ref{fig:op_dens} shows the \op{H2} in the desorbing gas, $R_{\rm thdes}(\ohh)/R_{\rm thdes}(\phh)$;
it is higher (i.e., the spin conversion probability upon adsorption becomes lower) with increasing the \ce{H2} gas density.
This trend does not depend on the parameter $f$, which is explained as follows.
In the model without thermal hopping, only sites with $\edcop \lesssim \edes \lesssim \edcdes$ efficiently contribute to the spin conversion.
At a given temperature, $\edcdes$ becomes smaller with increasing the \ce{H2} gas density, but $\edcop$ does not change (see Fig. \ref{fig:ecrit}); 
the number of sites that efficiently contribute to lowering the $\op{H2}$ in the desorbing gas becomes smaller with increasing the \ce{H2} gas density.
In the case with hopping, adsorbed \ce{H2} can visit various potential sites before it is desorbed, 
and $\theta_o/\theta_p$ is similar across the surface regardless of $\edes$ as discussed above.
Let us define the averaged desorption rate of \ce{H2} ($k_{\rm des}^{\rm av}$) on the surface, which should satisfy
\begin{align}
k_{\rm des}^{\rm av}\Theta N_{\rm site} = (1-\Theta)S\num{H_2}v_{\rm th}\sigma, \label{eq:kdes_tot}
\end{align}
under the adsorption-desorption equilibrium (see also the Appendix).
Then the averaged residence time of an adsorbed \ce{H2} on the surface, 1/$k_{\rm des}^{\rm av}$, is proportional to $\Theta/(1-\Theta)$ and inversely proportional to $\num{H_2}$.
As $\Theta$ depends only weakly on $\num{H_2}$ (Fig. \ref{fig:cov_eq}), 
the average residence time on the surface is reduced with increasing $\num{H_2}$.
Therefore, the probability of the ortho-para conversion upon adsorption is reduced with increasing $\num{H_2}$ in the case with hopping as well.

Finally, the spin conversion timescale of gaseous $\ohh$ ($\tau_{op}$) normalized by $\tau_{\rm col}$ is shown in the bottom panel of Figure \ref{fig:op_dens}.
The normalized conversion timescale ($\tau_{op}/\tau_{\rm col}$) corresponds to the average number of $\ohh$ collision to the surface to be required to produce one $\phh$. 
$\tau_{op}/\tau_{\rm col}$ increases with increasing $\num{H_2}$,
because the timescale for gaseous and solid \ce{H2} interaction becomes longer (the top panel) and 
the spin conversion upon adsorption becomes less efficient (the middle panel) with increasing $\num{H_2}$.
Note that $\tau_{\rm col}$ is inversely proportional to $\num{H_2}$ and thus $\tau_{\rm op}$ in fact drops with increasing $\num{H_2}$.
The absolute value of $\tau_{op}/\tau_{\rm col}$ depends on the efficiency of thermal hopping;
$\tau_{op}/\tau_{\rm col}$ in the models with ($f\leq0.5$) is smaller than that in the model without thermal hopping by a factor of a few.

\epsscale{0.8}
\epsscale{1.0}

\subsection{Temperature dependence}
\begin{figure}[ht!]
\plotone{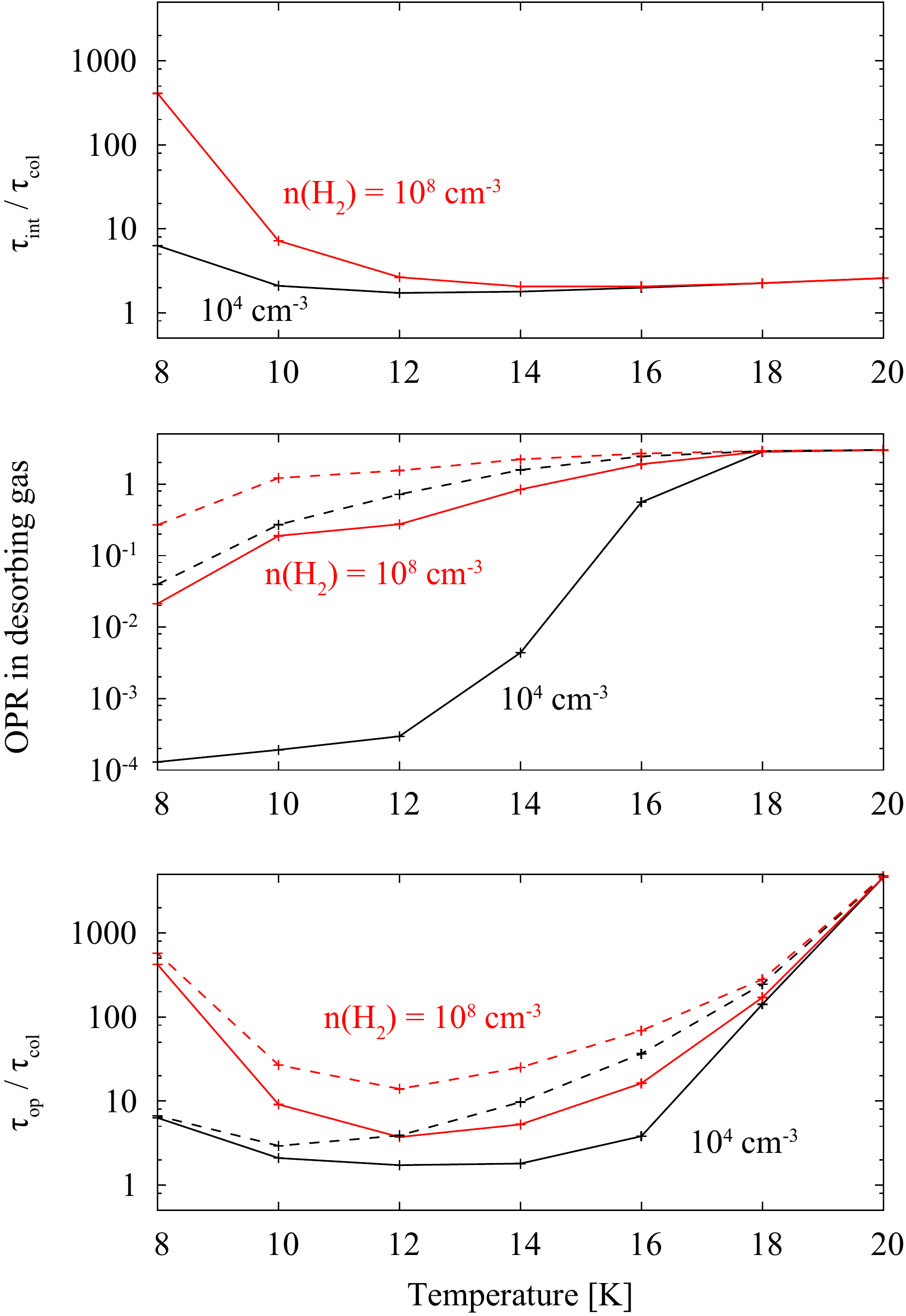}
\caption{Similar to Figure \ref{fig:op_dens}, but as functions of temperature. 
The \ce{H2} gas density is fixed to be either 10$^4$ cm$^{-3}$ (black) or 10$^8$ cm$^{-3}$ (red).
In the middle and bottom panels, solid lines represent the models with $f = 0.5$, 
while dashed lines represent the models without thermal hopping.}
\label{fig:op_tmpr}
\end{figure}

Figure \ref{fig:op_tmpr} is similar to Figure \ref{fig:op_dens}, but shows dependencies on temperature.
At given $\num{H_2}$, the normalized interaction timescale, $\tau_{\rm int}$/$\tau_{\rm col}$, becomes smaller with increasing temperature, 
because $\Theta$ decreases with increasing temperature.
On the other hand, the \op{H2} in the desorbing gas becomes higher (i.e., the spin conversion probability upon adsorption becomes lower) with increasing temperature.
As the temperature affects the interaction timescale and the conversion probability in the opposite direction, 
there is a critical temperature at which $\tau_{op}/\tau_{\rm col}$ is the smallest for given $\num{H_2}$:
$\sim$12-14 K for $\num{H_2} = 10^4$ cm$^{-3}$ and $\sim$10-14 K for $\num{H_2} = 10^8$ cm$^{-3}$.
At the lower temperatures, the exchange of gaseous and icy \ce{H2} is inefficient
(i.e., adsorbed \ce{H2} does not desorb and hinders another gaseous \ce{H2} to be adsorbed),
while at the higher temperatures, the residence time of \ce{H2} on surfaces is too short for the spin conversion.

\begin{figure}[ht!]
\plotone{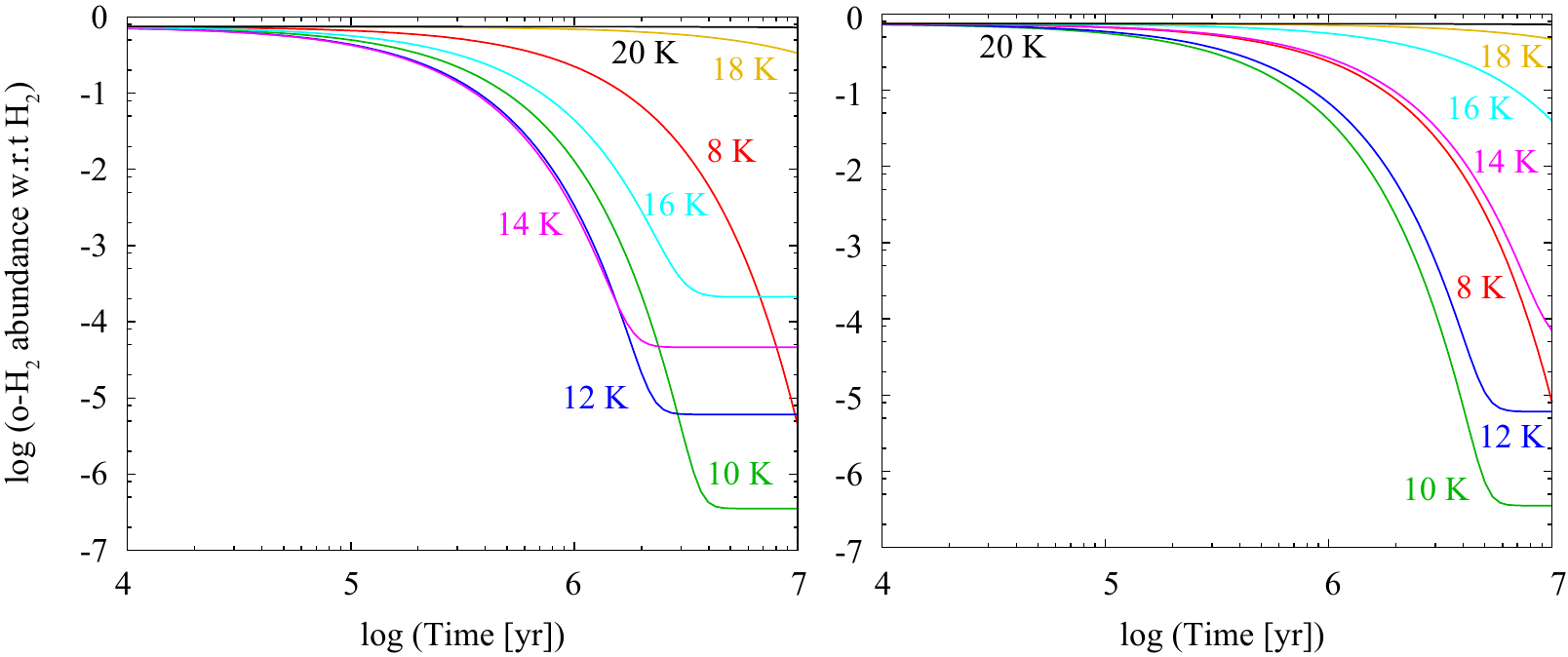}
\caption{Temporal evolution of the abundance of $\ohh$ in the gas phase with respect to \ce{H2} at $\num{H_2} = 10^4$ cm$^{-3}$, varying temperature from 8 K to 20 K 
in the models with the parameter $f$ of 0.5 (left panel) and in the models without thermal hopping (right panel).
}
\label{fig:h2opr}
\end{figure}

Figure \ref{fig:h2opr} shows the long term evolution of the $\ohh$ abundance in the gas phase with respect to \ce{H2} 
in the models with (left) and without thermal hopping (right) at $\num{H_2} = 10^4$ cm$^{-3}$, varying temperature from 8 K to 20 K.
Again, the temperature dependence of the spin conversion time scale is non-monotonic.
Note that the steady-state abundances depends on the temperature and are given by $9\exp(-170.5/T)$.

\epsscale{0.8}
\epsscale{1.0}

\epsscale{1.0}
\epsscale{1.0}

\section{Discussion}
\label{sec:discuss}
\subsection{A simple model for the \ce{H2} spin conversion rate}
For the spin conversion of \ce{H2} on grain surfaces in the ISM, the binding energy distribution and the thermal hopping among sites are essentially important.
In astrochemical simulations of star- and planet-forming regions, the rate-equation approach is usually employed
to describe the gas-phase and grain-surface chemistry.
In rate equation models, binding energy distribution is normally neglected and the binding energy of each species is represented as a single ``representative'' value. 
The goal of this subsection (and Appendix \ref{sec:appenda}) is to derive simple equations that reproduce our full numerical simulations of the overall  (gas + solid) $\op{H_2}$ evolution 
and can be implemented in existing astrochemical codes easily.

The spin conversion rate of overall \ce{H2} via the conversion on surfaces can be described as 
$\eta_{\rm op}(1-\Theta)SR_{\rm col}(\ohh)$ and $\eta_{\rm po}(1-\Theta)SR_{\rm col}(\phh)$, 
where $\eta_{\rm op}$ ($\eta_{\rm po}$) is the yield of gaseous $\phh$ ($\ohh$) per $\ohh$ ($\phh$) adsorption.
$\eta_{\rm op}(1-\Theta)S$ expresses the yield of $\phh$ per $\ohh$ collision to dust grain surfaces, which is equivalent to $\tau_{\rm col}/\tau_{op}$.
If such $\eta_{\rm op}$ and $\eta_{\rm po}$ are given, 
the time evolution of the \ce{H2} OPR in the gas phase via the spin conversion on grain surfaces can be obtained by solving simple rate equations, assuming the adsorption-desorption equilibrium of \ce{H2}:
\begin{align}
\frac{d\num{\ohh}}{dt} &= - \eta_{\rm op}(1-\Theta)SR_{\rm col}(\ohh) + \eta_{\rm po}(1-\Theta)SR_{\rm col}(\phh), \label{eq:simple_rate1} \\
\frac{d\num{\phh}}{dt} &= - \eta_{\rm po}(1-\Theta)SR_{\rm col}(\phh) + \eta_{\rm op}(1-\Theta)SR_{\rm col}(\ohh). \label{eq:simple_rate2}
\end{align}
Once the binding energy distribution of \ce{H2} is given, it is straightforward to calculate $\Theta$ using Eqs. \ref{eq:Thetat}, \ref{eq:fd_dist}, and \ref{eq:edcdes}.

The expression of $\eta_{\rm op}$ has been proposed by \citet{fukutani13} as
\begin{equation}
\eta_{\rm op}^{FS13} = \frac{k_{\rm op}^{\rm surf}}{k_{\rm op}^{\rm surf} + k_{\rm thdes}}, \label{eq:fk13}
\end{equation}
which describes the competition between the spin conversion and thermal desorption of adsorbed \ce{H2}.
It has been used in astrochemical simulations \citep{bovino17}.
While Eq. \ref{eq:fk13} is valid when surface property can be described by single binding energy, 
it does not take into account the binding energy distribution and the thermal hopping among various sites.
We develop more rigorous expression of $\eta_{\rm op}$ and $\eta_{\rm po}$, 
which reproduces our numerical results.
Our strategy is as follows: 
we first construct $\eta_{\rm op}$ and $\eta_{\rm po}$ that are adequate in two extreme cases, the fast hopping case and the slow (no) hopping case, 
and then combine the two extremes to obtain a general expression.
The derivation and formulations of $\eta_{\rm op}$ and $\eta_{\rm po}$ are described in the Appendix.

\begin{figure}[ht!]
\plotone{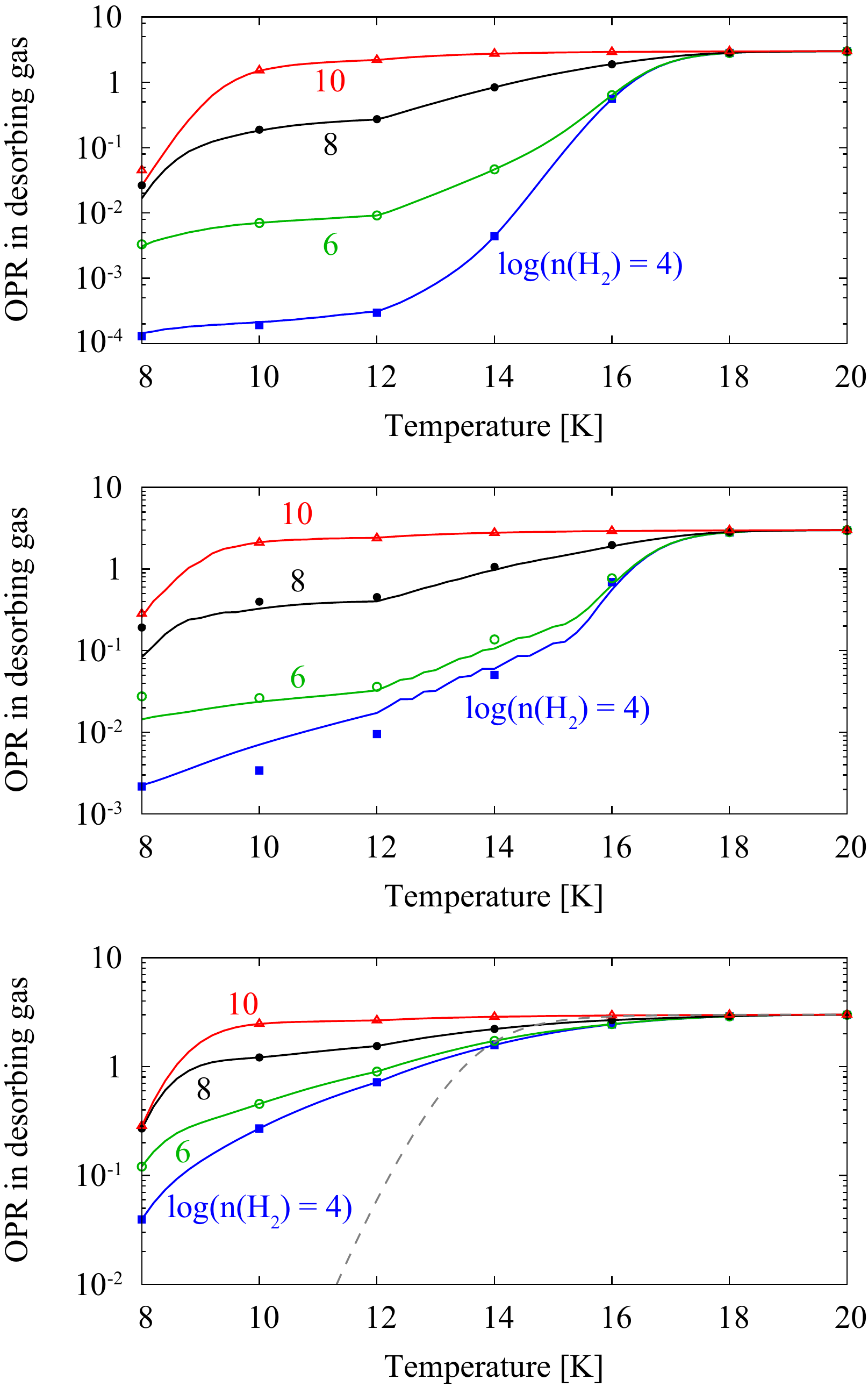}
\caption{$\op{H_2}$ in the desorbing gas estimated by Eq. \ref{eq:opsurf} with our  $\eta_{\rm op}$ and  $\eta_{\rm po}$  (lines) 
compared with that obtained by our numerical simulations (points) 
when $f=0.5$ (top panel), $f=0.8$ (middle panel), and thermal hopping is turned off (bottom panel).
$\op{H_2}$ in adsorbing \ce{H2} is set to be three.
\ce{H2} gas density is set to be 10$^4$ cm$^{-3}$ (blue), 10$^6$ cm$^{-3}$ (green), 10$^8$ cm$^{-3}$ (black), or 10$^{10}$ cm$^{-3}$ (red).
Gray dashed line in the bottom panel shows $\op{H_2}$ in the desorbing gas estimated by Eq. \ref{eq:opsurf} 
with Eq. \ref{eq:fk13} and with $k_{\rm thdes} = \nu \exp(-440/T)$.
}
\label{fig:opeff}
\end{figure}

Using $\eta_{\rm op}$ and $\eta_{\rm po}$, the $\op{H_2}$ of the desorbing gas from the surface can be expressed as 
\begin{align}
\frac{(1-\Theta)S[(1-\eta_{\rm op})f_{\rm o} + \eta_{\rm po}f_{\rm p}]}{(1-\Theta)S[\eta_{\rm op}f_{\rm o} + (1-\eta_{\rm po})f_{\rm p}]}, \label{eq:opsurf}
\end{align}
where $f_{\rm o}$ and $f_{\rm p}$ are the fraction of $\ohh$ and $\phh$, respectively, in adsorbing \ce{H2} (or equivalently \ce{H2} in the gas phase).
The factor $1-\eta_{\rm op}$ indicates the probability that adsorbed $\ohh$ desorbs as $\ohh$.
We set $f_{\rm o}$ and $f_{\rm p}$ to be 0.75 and 0.25, respectively, and compare Eq. \ref{eq:opsurf} with the numerical results 
(i.e., $R_{\rm thdes}(\ohh)/R_{\rm thdes}(\phh)$) at $t \sim10 (10^4$ cm$^{-3}$/$\num{H_2}$) yr or $t \sim 10^{-3}$ yr, whichever is longer) in Figure \ref{fig:opeff}.
In the case where the thermal hopping is efficient ($f \leq 0.5$) or where thermal hopping is turned off, 
the $\op{H_2}$ of the desorbing gas obtained by Eq. \ref{eq:opsurf} with our $\eta_{\rm op}$ and $\eta_{\rm po}$ 
almost perfectly agrees with the results of the full numerical simulation.
In the case of $f=0.8$, the two results are deviate, but only by a factor of two at most.
In the bottom panel of Figure \ref{fig:opeff},  Eq. \ref{eq:opsurf} evaluated with Eq. \ref{eq:fk13} and $k_{\rm thdes} = \nu \exp(-440/T)$ is also shown (gray dashed line).
We chose 440 K as ``representative'' binding energy of \ce{H2} \citep[e.g.,][]{cuppen07}.
In this case, the $\op{H_2}$ starts to sharply drop at $\sim$13 K, where $\edcop \sim 440$ K (see Fig. \ref{fig:ecrit}), because of the exponential dependence of $k_{\rm thdes}$.
The comparison clearly demonstrates that Eq. \ref{eq:fk13} does not reproduce our numerical results.

\epsscale{0.8}
\epsscale{1.0}

\subsection{Conversion on the surface versus in the gas phase}
So far, we investigated the efficiency of the \ce{H2} spin conversion on grain surfaces by solving the rate equations of gas-phase and grain-surface \ce{H2}.
Our detailed modeling has revealed that the efficiency of the \ce{H2} spin conversion on grain surfaces depends on 
the temperature, the \ce{H2} gas density, and the thermal hopping rates.
In the ISM, the $\op{H_2}$ is mostly determined by the competition between 
the \ce{H2} formation on surfaces and the spin conversion (i.e., thermalization) in the gas phase and that on surfaces.
In the dense ISM,  hydrogen is predominantly present in molecular form, and only small fraction of hydrogen is in atomic form, 
which is produced via a sequence of gas-phase reactions initiated by the cosmic-ray ionization of \ce{H2} \citep[e.g.,][]{tielens05}.
Atomic hydrogen can recombine on grain surfaces to reform \ce{H2}.
The OPR of \ce{H2} upon formation on surfaces is three, while the thermalized value of the OPR is on the order of 10$^{-7}$ at 10 K.
This significant deference makes the \ce{H2} formation important for the \ce{H2} OPR evolution in the dense ISM, 
even if the rate of \ce{H2} formation is much lower than the spin conversion rate \citep[see e.g.,][]{furuya18}.
Here we investigate the evolution of the $\op{H_2}$ in full gas-grain chemical reaction network model, 
in which the three relevant processes, the \ce{H2} formation and the spin conversion in the gas phase and on grain surfaces, are considered.
The main question we would like to explore here is at which conditions the spin conversion on grain surfaces dominates over the conversion in the gas phase. 

We run a grid of full gas-grain chemical reaction network model, which includes a variety of gaseous and icy species in addition to \ce{H2}.
The model is based on \citet{furuya15}, but additionally considers the spin conversion on grain surfaces using our $\eta_{op}$ and $\eta_{po}$.
In the model of \citet{furuya15}, the gas-ice chemistry is described by a three-phase model, in which three distinct phases, gas-phase, icy grain surface, 
and the bulk of ice mantle are considered \citep{hasegawa93}.
Gas-phase reactions, gas-surface interactions, and surface reactions are considered.
The chemical reaction network includes nuclear spin states of \ce{H2} and \ce{H+3} and deuterated species.
The \ce{H2} spin conversion in the gas phase through proton exchange reactions with \ce{H+} and with \ce{H3+} is included \citep{gerlich90,honvault11}. 
For this work, we exclude deuterated species for simplicity.

We run a grid of pseudo-time dependent models (i.e., the gas density and the temperature are fixed in each model), varying $\num{H_2}$ from 10$^4$ cm$^{-3}$ to 10$^8$ cm$^{-3}$ and temperature from 8 K to 20 K.
For each physical condition, we run three models, varying the treatment of  the \ce{H2} spin conversion on grain surfaces:
the model without the conversion on the surfaces, the model in which $\eta_{op}$ and $\eta_{po}$ are calculated assuming $f = 0.5$, and 
the model in which $\eta_{op}$ and $\eta_{po}$ are calculated neglecting thermal hopping.
$\Theta$ is calculated using Eqs. \ref{eq:Thetat} and \ref{eq:fd_dist}, and the sticking probability of \ce{H2} is taken from \citet{he16}.
We assume uniform grain radius of 0.1 $\mu$m with the dust-to-gas mass ratio of $10^{-2}$.
Elemental abundance ratios for H:He:C:N:O:Na:Mg:Si:S:Fe are 1.00:9.75(-2):7.86(-5):2.47(-5):1.80(-4):2.25(-9):1.09(-8):9.74(-9):9.14(-8):2.74(-9), where $a(-b)$ means $a\times10^{-b}$ \citep{aikawa99}.
Initially, all elements except for hydrogen is in atomic form, while hydrogen is present as \ce{H2} with the OPR of three.
The cosmic-ray ionization rate of \ce{H2} is set to be $\xi = 1.3\times10^{-17}$ s$^{-1}$.

\begin{figure}[ht!]
\plotone{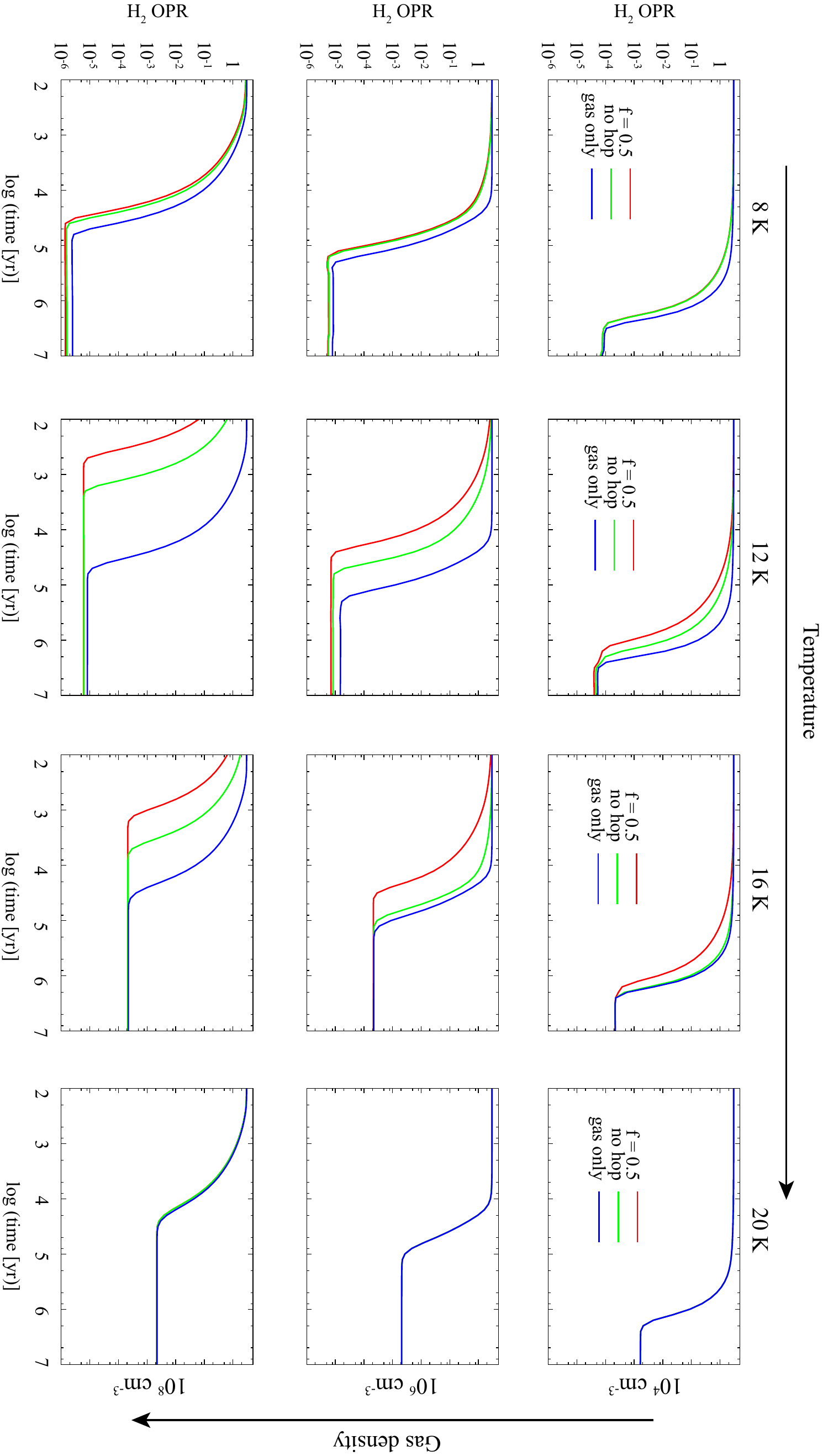}
\caption{Temporal variation of $\op{H2}$ in the gas phase in 
the models without the conversion on the surfaces (blue), the models in which $\eta_{op}$ and $\eta_{po}$ are calculated assuming $f = 0.5$ (red), and 
the models in which $\eta_{op}$ and $\eta_{po}$ are calculated neglecting thermal hopping (green), 
varying the $\num{H_2}$ gas density from 10$^4$ cm$^{-3}$ to 10$^8$ cm$^{-3}$ and temperature from 8 K to 20 K. 
}
\label{fig:h2opr_full}
\end{figure}

Figure \ref{fig:h2opr_full} shows the temporal variation of the $\op{H2}$ in the gas phase.
The steady-state value of the $\op{H2}$ in our models is higher than the thermalized value at $\lesssim$16 K due to the \ce{H2} formation on grain surfaces \citep[e.g.,][]{furuya18}.
We find that the conversion on the surfaces dominates over that in the gas phase at the temperatures below 20 K, 
regardless of the \ce{H2} gas density and the thermal hopping rates of \ce{H2}.
The rate of the spin conversion on the surface drops at temperatures higher than the critical temperature, while that in the gas phase is not sensitive to the temperature in the range of 8 K to 20 K.
The impact of the spin conversion on the surfaces becomes more significant with increasing the \ce{H2} gas density;
the timescale of the spin conversion via the gas phase proton exchange reactions roughly scales with $(\num{H_2})^{-0.5}$, 
while the collisional timescale of \ce{H2} to the surface scales with $(\num{H_2})^{-1}$ and $\eta$ depends only weakly on $\num{H_2}$ 
(see the bottom panel of Fig. \ref{fig:op_dens}).
The rate of spin conversion in the gas phase and that on the grain surfaces depend differently on physical and chemical conditions ($T, \num{H_2}, \xi$, $R_{\rm col}(\ce{H2})$, etc.).
Therefore, one has to consider the spin conversion both in the gas phase and on grain surfaces for accurate modeling of the $\op{H2}$ evolution in star- and planet forming regions, 
which cover wide ranges of the physical and chemical conditions.

Our model was constructed using the experimentally derived binding energy distribution of \ce{H2} 
and the nuclear spin conversion rate on water ice surfaces.
In star- and planet-forming regions, gaseous \ce{H2} would interact with not only water ice surfaces, but also various types of surfaces, including silicates, graphites, and CO ices.
\ce{H2} molecules should be formed well before dust grains are coated by water ice mantles.
Infrared ice observations have found that the catastrophic CO freeze out happens in dense cores, and ice layers, which mainly consist of CO and \ce{CH3OH} 
are formed on top of the water ice layers \citep[e.g.,][]{pontoppidan06}.
To the best of our knowledge, similar experimental measurements adequate for bare dust grains and ices other than water are not available in the literature.
Once such measurements become available, it is straightforward to apply our models to the other types of surfaces, 
and to simulate the evolution of the \op{H2} from the formation stage of molecular clouds to the dense core stage \citep[e.g.,][]{furuya15}, 
considering the \ce{H2} spin conversion both on surfaces and in the gas phase.
This is the necessary step for better understanding of the \op{H2} especially in the early stages of star formation.

\epsscale{0.8}
\epsscale{1.0}

\section{Conclusion}
\label{sec:conclusion}
The ortho-to-para ratio of \ce{H2} can significantly affect the molecular evolution, for example deuterium fractionation, in the ISM.
The main mechanism of the \ce{H2} ortho-para conversion, i.e., whether in the gas phase or on grain surfaces, remains unclear, 
because the efficiency of the latter in the ISM is not well understood.
In this work, we have studied the impact of the nuclear spin conversion of \ce{H2} on the water ice surface 
on the evolution of the overall (gas+ice) \ce{H2} under the physical conditions 
that are relevant to star- and planet-forming regions.
We have constructed the rate equation model that considers adsortption of gaseous \ce{H2}, 
thermal hopping, desorption, and the nuclear spin conversion of adsorbed \ce{H2}.
We have used the experimentally derived binding energy distribution of \ce{H2} and the nuclear spin conversion rate on amorphous water ice surfaces.
It was found that the spin conversion efficiency depends on \ce{H2} gas density and surface temperature. 
There are two main factors that control the efficiency of the spin conversion:
(i) the efficiency of gaseous and solid \ce{H2} interaction and 
(ii) the probability of the spin-state conversion of an adsorbed \ce{H2} molecule before it is desorbed.
Enhanced \ce{H2} gas density reduces the spin conversion efficiency, because the \ce{H2} coverage on the surface increases with increasing the \ce{H2} gas density, 
which hinders gaseous \ce{H2} molecules to be adsorbed.
The temperature dependence is not monotonic; 
there is a critical surface temperature at which the efficiency is the maximum.
At low temperatures, the exchange of gaseous and icy \ce{H2} is inefficient,
while at warm temperatures, the residence time of \ce{H2} on surfaces is too short for the spin conversion.

By constructing the full gas-ice chemistry model with the \ce{H2} spin conversion on grain surfaces,
we have found that the spin conversion on the surface dominates over that in the gas-phase at the temperatures below 20 K, 
regardless of the \ce{H2} gas density and the thermal hopping rate of \ce{H2} on the surface.
We have developed a simple, but accurate formulation to implement the nuclear spin conversion on grain surfaces 
in existing gas-ice astrochemical models (see Appendix).
Our formulation can be applied to any other types of surfaces (e.g., bare dust grain surfaces and CO ice surfaces), 
once the nuclear spin conversion rate, the sticking probability, and the binding energy distribution of \ce{H2} on the other surfaces become available.

\acknowledgments
We thank the anonymous referee for the useful comments.
This work is partly supported by JSPS KAKENHI Grant numbers, 17K14245 and 17H06087.


\begin{appendix}
\section{Construction of parameter $\eta$}
\label{sec:appenda}
Here we derive equations for the yield of gaseous $\phh$ per $\ohh$ adsorption ($\eta_{\rm op}$) and the yield of gaseous $\ohh$ per $\phh$ adsorption ($\eta_{\rm po}$).
Our strategy is as follows.
We first construct $\eta_{\rm op}$ and $\eta_{\rm po}$ that are adequate in two extreme cases:
the fast hopping case, where thermal hopping of adsorbed \ce{H2} is fast and the \ce{H2} OPR is the same across the surface (Eqs. \ref{eq:etaf_op} and \ref{eq:etaf_po}) 
and the slow hopping case, where thermal hopping is negligible (Eq. \ref{eq:etas_op}).
Then we combine the two extremes to obtain a general expression (Eq. \ref{eq:eta_final}).

\subsection{In the limit of fast hopping}
As discussed in Sect. \ref{sec:result}, when thermal hopping is considered, $k_{\rm thdes}^{-1}$ is not a good measure for the residence time, but $(k_{\rm des}^{\rm av})^{-1}$ is.
We denote the conditional probability that the spin conversion on a surface from $\ohh$ to $\phh$ occurs before $\ohh$ desorption, and then $\phh$ desorption occurs before reconversion to $\ohh$ as $p_{\rm op}$:
\begin{align}
p_{\rm op} = \frac{k_{\rm op}^{\rm surf}}{k_{\rm op}^{\rm surf} + k_{\rm des}^{\rm av}}\cdot \frac{k_{\rm des}^{\rm av}}{k_{\rm po}^{\rm surf} + k_{\rm des}^{\rm av}}, \label{eq:p_op}
\end{align}
We denote the conditional probability that the spin conversion on a surface from $\ohh$ to $\phh$ occurs before $\ohh$ desorption, 
and then reconversion from $\phh$ to $\ohh$ occurs before $\phh$ desorption as $r$:
\begin{align}
r = \frac{k_{\rm op}^{\rm surf}}{k_{\rm op}^{\rm surf} + k_{\rm des}^{\rm av}}\cdot \frac{k_{\rm po}^{\rm surf}}{k_{\rm po}^{\rm surf} + k_{\rm des}^{\rm av}}.
\end{align}
Then, for example, $rp_{\rm op}$ is the probability that a sequence of spin conversion, $\ohh$ $\rightarrow$ $\phh$ $\rightarrow$ $\ohh$ $\rightarrow$ $\phh$, occurs on a surface and then $\phh$ desorbs.
Using $p_{\rm op}$ and $r$, the yield of gaseous $\phh$ per $\ohh$ adsorption in the limit of fast hopping ($\eta^{(f)}$) may be written as follows:
\begin{align}
\eta^{(f)}_{\rm op} &= p_{\rm op} +  rp_{\rm op} +  r^2p_{\rm op} + r^3p_{\rm op} +... \label{eq:etaf_op} \\ 
                   &= p_{\rm op}/(1-r), \nonumber \\
                   &= \frac{k^{\rm surf}_{\rm conv}}{k^{\rm surf}_{\rm conv} + k_{\rm des}^{\rm av}} \cdot \frac{1}{1+\gamma}, \nonumber
\end{align}
where $k^{\rm surf}_{\rm conv} =1/\tau^{\rm surf}_{\rm conv}$.
Note that $\eta_{\rm op}$ given above considers the possibility of multiple spin conversion on a surface.
Similarly, $\eta^{(f)}_{\rm po}$ is given by
\begin{align}
\eta^{(f)}_{\rm po} = \frac{k^{\rm surf}_{\rm conv}}{k^{\rm surf}_{\rm conv} + k_{\rm des}^{\rm av}} \cdot \frac{\gamma}{1+\gamma}. \label{eq:etaf_po}
\end{align}

\subsection{In the limit of slow hopping}
In the limit of slow hopping, we can treat sites with a different energy depth separately.
Then we define $\eta^{(s)}$ for each $\edes$ using $k_{\rm thdes}$ as follows:
\begin{align}
\eta^{(s)}_{\rm op}(\edes) &= \frac{k^{\rm surf}_{\rm conv}}{k^{\rm surf}_{\rm conv} + k_{\rm thdes}(\edes)} \cdot \frac{1}{1+\gamma}, \\
\eta^{(s)}_{\rm po}(\edes) &= \frac{k^{\rm surf}_{\rm conv}}{k^{\rm surf}_{\rm conv} + k_{\rm thdes}(\edes)} \cdot \frac{\gamma}{1+\gamma}. 
\end{align}
Again $\eta^{(s)}(\edes)$ considers multiple spin conversion on a surface, in contrast to Eq. \ref{eq:fk13}, where only single spin conversion is considered.

We define $\langle\eta^{(s)}_{\alpha\beta}\rangle$, where $\alpha$ and $\beta$ are $o$ or $p$, as the average of $\eta^{(s)}_{\alpha\beta}(\edes)$ weighted by thermal desorption rates;
\begin{align}
\langle\eta^{(s)}_{\alpha\beta}\rangle &= \int \eta^{(s)}_{\rm \alpha\beta}(\edes')k_{\rm thdes}(\edes')\theta(\edes')g(\edes')d\edes' \bigg/ \int k_{\rm thdes}(\edes')\theta(\edes')g(\edes')d\edes', \label{eq:etas_op} 
\end{align}
where the integration range is from $\edes^{\rm thresh}$ (defined below) to $\infty$.

\subsection{General case}
\label{sec:append3}
We denote the threshold binding energy as $\edes^{\rm thresh}$; 
sites with the binding energy lower (higher) than $\edes^{\rm thresh}$ is considered in the fast (slow) hopping regime.
$\edes^{\rm thresh}$ is defined as the binding energy that satisfies
\begin{align}
k^{\rm surf}_{\rm conv} = [1-\theta(\edcop)]k_{\rm hop}(\edes^{\rm thresh} \rightarrow \edcop),
\end{align}
and $\edes^{\rm thresh} > \edcop$.
Note that $k_{\rm hop}(\edes^{\rm thresh} \rightarrow \edcop)$ is the lower limit of $k_{\rm hop}(\edes \rightarrow \edes')$ 
where $\edcop < \edes < \edes^{\rm thresh}$ and $\edcop < \edes' < \edes^{\rm thresh}$ (Eq. \ref{eq:ehop}).
For sites with $\edcop < \edes < \edes^{\rm thresh}$, the thermal hopping rate is greater than the spin conversion rate, and the thermal desorption rate is smaller than the two rates 
(i.e., $k_{\rm thdes} < k^{\rm surf}_{\rm conv} < k_{\rm hop}$).
Thus the $\op{H_2}$ in such sites are expected to be similar (see the middle panel of Fig. \ref{fig:h2op_cov}).
At $\num{H_2} = 10^4$ cm$^{-3}$, for example, $\edes^{\rm thresh}$ is 596 K for $f=0.5$ and 393 K for $f=0.8$.
In sites with $\edes < \edcop$, thermal desorption is more efficient than the spin conversion.
Thus the $\op{H_2}$ in sites with $\edes < \edcop$ is mostly determined by the competition between adsorption and thermal hopping from sites with $\edes > \edcop$.
For simplicity, we assume that the $\op{H_2}$ in sites with $\edes < \edcop$ is the same as that in sites with  $\edcop < \edes < \edes^{\rm thresh}$.

Using $\edes^{\rm thresh}$, we define $\Theta^{(f)}_{\rm max}$, $\Theta^{(s)}_{\rm max}$, $\Theta^{(f)}$, and $\Theta^{(s)}$ as
\begin{align}
\Theta^{(f)}_{\rm max} = \int_{0}^{\edes^{\rm thresh}}g(\edes')d\edes', \,\,\, 
\Theta^{(s)}_{\rm max} = \int_{\edes^{\rm thresh}}^{\infty}g(\edes')d\edes',\\
\Theta^{(f)} = \int_{0}^{\edes^{\rm thresh}}\theta(\edes')g(\edes')d\edes', \,\,\, 
\Theta^{(s)} = \int_{\edes^{\rm thresh}}^{\infty}\theta(\edes')g(\edes')d\edes',
\end{align}
where $\Theta^{(f)}_{\rm max}$ ($\Theta^{(s)}_{\rm max}$) is the fraction of sites that is considered in the fast (slow) hopping regime.
$\Theta^{(f)}$ ($\Theta^{(s)}$) is a subset of the \ce{H2} coverage that is considered in the fast (slow) hopping regime.
Note that $\Theta^{(f)}_{\rm max} + \Theta^{(s)}_{\rm max} = 1$ and $\Theta^{(f)} + \Theta^{(s)} = \Theta$.

Using $\eta^{(f)}$ and $\langle\eta^{(s)}\rangle$ defined above, we define the general expression of the spin conversion yield upon adsorption, $\eta$, as
\begin{align}
\eta_{\rm \alpha\beta} = \frac{\Theta^{(f)}_{\rm max} - \Theta^{(f)}}{1-\Theta}\eta_{\alpha\beta}^{(f)} + \frac{\Theta^{(s)}_{\rm max} - \Theta^{(s)}}{1-\Theta}\langle \eta_{\alpha\beta}^{(s)} \rangle. \label{eq:eta_final}
\end{align}
Note that $\eta_{\rm po}/\eta_{\rm op}=\gamma$, because $\eta^{(f)}_{\rm po}/\eta^{(f)}_{\rm op}=\gamma$ and $\langle \eta^{(s)}_{\rm po} \rangle/\langle \eta^{(s)}_{\rm op} \rangle=\gamma$.
We realized that in the evaluation of $k_{\rm des}^{\rm av}$, using $\Theta^{(f)}$ is more reasonable rather than using $\Theta$.
Then we redefine $k_{\rm des}^{\rm av}$ as
\begin{align}
k_{\rm des}^{\rm av}\Theta^{(f)} N_{\rm site} = (\Theta^{(f)}_{\rm max}-\Theta^{(f)})S\num{H_2}v_{\rm th}\sigma,
\end{align}
and use this in the evaluation of $\eta^{(f)}$.
$\eta_{\rm op}$ calculated by Eq. \ref{eq:eta_final} is listed in Table \ref{table:eta}.

\begin{table}
\begin{center}
\caption{$\eta_{\rm op}$ evaluated by Equation \ref{eq:eta_final} \label{table:eta}}
\begin{tabular}{cccccccc}
\hline\hline
     &                             \multicolumn{3}{c}{$f=0.5$}                            &     &      \multicolumn{3}{c}{w/o thermal hopping}                            \\   
\cline{2-4} \cline{6-8}
     & 10$^{4}$ cm$^{-3}$   & 10$^{6}$ cm$^{-3}$  & 10$^{8}$ cm$^{-3}$ &     &10$^{4}$ cm$^{-3}$   & 10$^{6}$ cm$^{-3}$  & 10$^{8}$ cm$^{-3}$   \\ 
\hline
8 K &  9.998(-1)               & 9.959(-1)               &  9.781(-1)             &      &  9.493(-1)       &  8.563(-1)                &     7.181(-1)     \\
10 K & 9.972(-1)               & 9.906(-1)               &  7.948(-1)            &       &  7.162(-1)       &  5.839(-1)               &      2.687(-1)    \\
12 K & 9.959(-1)               & 9.879(-1)               &   7.164(-1)           &       &  4.427(-1)       &  3.687(-1)                &     1.907(-1)      \\
14 K & 9.942(-1)               & 9.409(-1)               &    3.912(-1)          &       &   1.841(-1)      & 1.568(-1)                &      8.188(-2)       \\
16 K & 5.227(-1)                & 4.815(-1)               &    1.266(-1)         &       &    5.443(-2)     &  5.256(-2)               &      2.984(-2)       \\
18 K & 1.590(-2)                & 1.586(-2)               &     1.328(-2)        &       &     9.187(-3)    &  9.173(-3)               &      8.052(-3)       \\
20 K & 5.646(-4)                & 5.646(-4)                &     5.613(-4)       &       &     5.484(-4)    &  5.484(-4)               &      5.453(-4)       \\
\hline
\end{tabular}
\end{center}
\tablecomments{$a(-b)$ means $a\times 10^{-b}$.}
\end{table}

\end{appendix}




\end{document}